\documentclass[12pt]{article}
\usepackage[T2A]{fontenc}
\usepackage[utf8]{inputenc}
\usepackage{tikz}
\usepackage[margin=.75in]{geometry}
\usepackage[margin=.5in]{caption}
\usepackage{
amsmath,
amssymb,
amsthm,
amsrefs,
mathrsfs,
hyperref,
xcolor,
enumitem 
}

\usepackage{listings}
\definecolor{dkblue}{rgb}{0,0.1,0.5}
\definecolor{lightblue}{rgb}{0,0.5,0.5}
\definecolor{dkgreen}{rgb}{0,0.4,0}
\definecolor{lightgreen}{rgb}{0,0.3,0}
\definecolor{dk2green}{rgb}{0.4,0,0}
\definecolor{dkviolet}{rgb}{0.8,0,0.1}

\usepackage[para]{footmisc}

\newcommand{\Coq}{{\sc Coq }}
\newcommand{\MC}{{\sc Mathematical Components }}
\newcommand{\ssr}{{\sc SSReflect }}

\newcommand{\N}{\mathbb{N}}

\newtheorem{theorem}{Theorem}

\newtheorem{lemma}[theorem]{Lemma}

\makeatletter
\renewcommand{\section}{\@startsection
{section}
{1}
{0mm}
{-2\baselineskip}
{1\baselineskip}
{\normalfont\large\scshape\centering}} 
\makeatother

\makeatletter
\renewcommand{\subsection}{\@startsection
{subsection}
{2}
{0mm}
{-\baselineskip}
{1 \baselineskip}
{\normalfont\scshape}} 
\makeatother

\makeatletter
\renewcommand{\subsubsection}{\@startsection{subsubsection}{3}{\z@}%
  {3.25ex \@plus 1ex \@minus .2ex}{-1em}{\normalfont\normalsize\itshape}}
\makeatother

\title{\textbf{{\scshape \large Formal verification of Zagier's one-sentence proof}}}

\author{Guillaume Dubach\footnote{guillaume.dubach@ist.ac.at} \ \& Fabian Mühlböck\footnote{fabian.muehlboeck@ist.ac.at} \\ IST Austria}

\date{}

\begin{document}

\maketitle

\hfill \small{\textit{It is futile to do with more things that which can be done with fewer.} \quad }

\hfill William of Ockham, \textit{Summa Logicae} i.12 \quad \quad

\vspace{.2in}
\abstract{
We comment on two formal proofs of Fermat's sum of two squares theorem, written using the \MC libraries of the \Coq proof assistant. The first one follows Zagier's celebrated one-sentence proof; the second follows David Christopher's more recent proof relying on partition-theoretic arguments. Both formal proofs rely on a general property of involutions of finite sets, of independent interest.
The proof technique consists for the most part of automating recurrent tasks (such as case distinctions and computations on natural numbers) via ad hoc tactics. With the same method, we also provide a formal proof of another classical result on primes of the form $a^2 + 2 b^2$.
}

\vspace{-.1in}
\section*{Introduction}
In his 1990 paper \cite{Zagier1990}, Zagier provided what he advertised as a `one-sentence proof' of the following proposition, first stated by Fermat in a letter to Mersenne:

\begin{theorem}[Fermat]\label{Fermat_thm}
Every prime number $p$ such that $p \equiv 1 \ \mathrm{mod} \ 4$ is a sum of two squares, i.e. there exist $a,b \in \mathbb{N}$ such that $p=a^2+b^2$.
\end{theorem}

Zagier's proof relies on the basic fact that an involution $f$ on a finite set $S$ is such that the number of fixed points of $f$ and the cardinality $|S|$ have the same parity. This principle is used twice, in complementary ways: after defining an appropriate set $S$, a first involution is invoked to justify that $|S|$ is odd; this implies that \textit{another} involution on $S$ has at least one fixed point; and any such fixed point then yields two integers $a$ and $b$ solutions of the problem. The second involution is a very natural one, whereas the first involution appears at first sight to be extremely involved. The trick, in order to turn this argument into a `one-sentence proof', is to only state the definitions and major steps as one long sentence, leaving all details and especially all ancillary (but logically necessary) verification steps to the reader. This condensed writing style and the mysterious origin of the first involution undoubtedly endowed this proof with rare appeal when it first appeared, and still does some thirty years later. In particular, it prompted several generalizations and applications to other results of the same kind, a review of which is given in \cite{Elsholtz_review}; our method applies to these with very few changes. To illustrate this, we provide a formal proof of the following statement, following an argument first published by Jackson \cite{Jackson}, which was found independently by Elsholtz \cite{Elsholtz1} and Generalov \cite{Generalov}.

\begin{theorem}[Jackson]\label{Jackson_thm}
For every prime number $p$ such that $p \equiv \ 3 \ \mathrm{mod} \ 8$, there exist $a, b \in \N$ such that $p=a^2 + 2 b^2$.
\end{theorem}


More recently, another elegant argument for Theorem \ref{Fermat_thm} was given by David Christopher \cite{DavidChristopher2016}. His proof involves a few more steps, and the definition of four different finite sets instead of one, corresponding to partitions of the integer $p$ with different properties. The basic mechanism, as for Zagier's proof, is to consider involutions on these finite sets. The whole argument can hardly be stated as only one sentence, but it is safe to say that it takes less time to most readers to be convinced of its validity, as each step left to the reader is truly more elementary. \medskip

The purpose of this paper is to present these proofs in details and to comment on their formalisation using the \Coq proof assistant. The current state of the corresponding code can be found on the following GitHub repository:

\centerline{\url{https://github.com/gdubach/Zagier_project}.}

The code essentially relies on the \MC libraries, presented in the fundamental book \cite{MC_book}, and takes advantage of the \ssr (\textit{small scale reflection}) syntax, whose general presentation \cite{ssr_manual} is also included in the \Coq reference manual \cite{coq_reference}. The tutorial \cite{ssr_tuto} is another valuable reference on this particular syntax. \medskip

The general property of involutions over finite sets needed in all proofs is briefly presented in Section \ref{involution_section}, and might be of independent interest. Section \ref{Zagier_section} presents Zagier's original proof of Theorem \ref{Fermat_thm} together with a recent geometric interpretation of it; we also comment on the code methodology and how it applies to Theorem \ref{Jackson_thm} as well. Section \ref{Partition_section} similarly provides a general presentation of David Christopher's proof of Theorem \ref{Fermat_thm}, and comments on the features of the code that are specific to this alternative argument. \medskip

A general aspect of the formal verification of these proofs is that it somehow reverses the usual perspective, in the sense that the difficulties and subtleties concern mostly things that are self-evident for a human mind, and vice-versa: what takes some time to a mathematician, such as checking that Zagier's first map is indeed an involution, is effortless once the computation is automated. However, as far as the logical structure is concerned, the formal proof faithfully reflects the steps of the original proof. \medskip

Fermat's sum of two squares theorem already appears in the \Coq literature: it has been verified by Laurent Théry \cite{Thery2016}. Théry's formal proof relies on the properties of the ring $\mathbf{Z}[i]$ of Gauss integers -- a very different approach than the combinatorial arguments formalized here. 
Establishing all relevant properties of Gauss integers takes more lines of code than following Zagier's legerdemain approach, but it is undoubtedly a more robust and canonical technique. Other proofs of Fermat's theorem and related results have been written in other proof assistants as well; notably, the recent work \cite{JChanHOL} in \textsc{HOL4} also follows Zagier's proof.

\vspace{-.2in}
\section*{Acknowledgments}
We would like to thank Marie Kerjean for her help and advice with the \MC libraries, and Christian Elsholtz for pointing out the generalizations of Zagier's proof technique. Answers and advice from the \Coq community on Zulip \cite{zulip_chat} have been also of great help; special thanks to Cyril Cohen and Christian Doczkal for explaining to us some aspects of the \texttt{finmap} library.  \medskip

\noindent G. Dubach gratefully acknowledge funding from the European Union's Horizon 2020 research and innovation programme under the Marie Sk{\l}odowska-Curie Grant Agreement No. 754411. F. Mühlböck's research is supported in part by the Austrian Science Fund (FWF) under grant Z211-N23 (Wittgenstein Award).

\newpage
\section{Involutions of finite sets}\label{involution_section}

The proofs we formalize rely on a general property of involutions over finite sets:
\begin{lemma}\label{FLI}
If $S$ is a finite set, an involution $f: S \rightarrow S$ is such that
\begin{equation}
    \left| \{ s \in S \ : \ f(s)=s \} \right| \equiv |S| \ \mathrm{mod} \ 2.
\end{equation}
In particular, if $|S|$ is odd, every such involution has at least one fixed point.
\end{lemma}
In professional mathematical papers, this is typically a fact considered so intuitive, and indeed elementary, that a justification would be superfluous. Formalizing it, however, still demands careful consideration. To do so, we work in the context of a variable Type $K$ (which will be $\N^3$ or $\N^4$ in the proofs). It matters to the structure of the proof and especially to the partial automation that $K$ should be a possibly infinite type on which a function $f$ is defined, that is an involution only on a finite subset $E \subset K$. The notion of a \textit{finite} subset of a possibly \textit{infinite} type is given by the \texttt{fset} structure from the \texttt{finmap} folder, a recent and valuable addition to the \MC library. So we write, with the appropriate imports:
\begin{lstlisting}
Variable K: choiceType.

Definition involution_on (E: {fset K}) (f:K->K) := 
(forall x, x \in E -> f x \in E) /\ (forall x, x \in E -> f (f x) = x).

Definition fixedset (E: {fset K}) (f:K->K):{fset K} := [fset x in E | x == f x].
\end{lstlisting}
And the lemma we want to prove is stated as follows:
\begin{lstlisting}
Lemma involution_lemma (f:K->K): forall (E:{fset K}),
involution_on E f -> odd #|`(fixedset E f)| = odd #|`E|.
\end{lstlisting}

The proof is rather straightforward after a few technical lemmas. We perform a strong induction on the number of the non-fixed points of $f$, i.e. on the natural integer
\begin{verbatim}
n := #|`E`\`(fixedset E f)| : nat.
\end{verbatim}
The reason why strong induction is appropriate here is that a non-fixed point $x \neq f(x)$ can always be removed from the set $E$ together with its image $f(x)$, $f$ being still an involution on the smaller set $F_x:=E \backslash \{x,f(x)\}$. So the basic induction step reduces the number of non-fixed points from $n+2$ to $n$. Writing this as a strong induction also eliminates the need to treat the case $n=1$ explicitly; the fact that $n=1$ can never occur is embedded in the structure of the proof. \medskip

One technical aspect is that we need the induction hypothesis to state that the conclusion holds for \textit{any finite set} $E$ as well as any $f$ such that the number of non-fixed points of $f$ on $E$ is $n$. Indeed, we then want to apply it the same function $f$ restricted to the smaller set $F_x$. To conclude the proof, it is to be verified that
\begin{enumerate}[label=(\roman*)]
    \item $f$ is still an involution of $F_x$, 
    \item the number of fixed points did not change,
    \item the cardinality of the base set ($F_x$ instead of $E$) has decreased by $2$, and
    \item so has the cardinality of the non-fixed points.
\end{enumerate}

A shorter proof would certainly have been obtained by assuming the Type $K$ to be endowed with a total order (which can easily be done on $\N^3, \N^4$) and considering the subsets of $E$ such that $f(x)<x$, $f(x)=x$ and $f(x)>x$. We chose not to rely on an order structure for this proof, so that Lemma \ref{FLI} could be used in a less cumbersome way on any type $K$.

\newpage
\section{Zagier's one-sentence proof}\label{Zagier_section}

\subsection{The 1990 proof and a recent illustration}

Let us first re-state Zagier's proof the way he wrote it.

\begin{proof}[Zagier's one-sentence proof]
The involution on the finite set
\begin{equation}\label{set_Sp}
S_p := \{ (x,y,z)\in \mathbb{N}^3 \ | \ x^2+4yz=p \}
\end{equation}
defined by
\begin{equation}\label{zag_involution}
    (x,y,z) \mapsto
    \left\{
    \begin{array}{ll}
        (x+2z, z, y-x-z) & \text{if } x < y-z \\
        (2y-x, y, x-y+z) &  \text{if } y-z < x < 2y \\
        (x-2y, x-y+z, y) & \text{if } 2y < x
    \end{array}
    \right.
\end{equation}
has exactly one fixed point, so $|S_p|$ is odd, and the involution defined by 
\begin{equation}\label{zig_involution}
    (x,y,z) \mapsto (x,z,y).
\end{equation}
also has a fixed point.
\end{proof}

Writing the above proof in only one sentence is clearly a bravura piece; let us make a few comments to unpack the argument. It is first implicitly left to the reader to check
\begin{enumerate}[label=(\roman*)]
\item that (\ref{set_Sp}) defines a finite set $S_p$;
\item that (\ref{zag_involution}) defines an involution on $S_p$, which is far from obvious;
\item that (\ref{zag_involution}) has exactly one fixed point, which also requires to be worked out.
\end{enumerate}
From this, the elementary Lemma \ref{FLI} is invoked to deduce that $|S_p|$ is odd, which is the turning point of the proof. It is then, again, left to the reader to check that 
\begin{enumerate}[label=(\roman*)]
\item[(iv)] (\ref{zig_involution}) also defines an involution on $S_p$.
\end{enumerate}
Therefore, by a second appeal to Lemma \ref{FLI}, this involution on a finite set of odd cardinality must have at least one fixed point. By definition, such a fixed point is a triple $(x,y,y)\in S_p$, which implies that $p=x^2+4y^2$, a sum of two squares, which concludes the proof. \medskip

Concerning the general outline of the argument, Zagier refers to a former proof by Heath-Brown \cite{HeathBrown1984} and the works of Liouville. However, Zagier's proof contains a new ingredient, namely the involution defined by (\ref{zag_involution}), which constitutes the most intriguing part of his very short paper. It is not clear at first sight how anyone would come up with the idea of such a map, nor is it clear why this should be an involution on $S_p$. \medskip

In recent years, a geometric interpretation has become popular, and now appears in various forms online. Its origin seems to be a paper by A. Spivak \cite{Spivak2008}. It is a simple and beautiful idea that makes the involution (\ref{zag_involution}) and its distinction of three cases appear naturally. First, each triple $(x,y,z)$ is associated to a `windmill', i.e. a square of side $x$ surrounded by four rectangles of dimensions $y \times z$; it is important to specify that $y$ is the length of the side situated \textit{along} the $x \times x$ square. This being posed, the windmill is unique up to symmetry -- symmetric windmills being considered equivalent for the sake of the argument\footnote{For instance, Type 2 and Type 4 in Figure \ref{5_types} do not have the same orientation, but what matters is their common outer shape, as explained below.}. Such windmills can be decomposed in five types, according to the relative size of $y$ compared to $x$ and $x+z$.

\newpage

\begin{figure}[t!]
$$
\begin{array}{|c|c|c|c|c|}
\hline
\text{Type } 1 & \text{Type } 2 & \text{Type } 3 & \text{Type } 4 & \text{Type } 5 \\
\hline
y<x/2 & x/2<y<x & x=y & x<y<x+z & x+z<y \\
\hline
\begin{tikzpicture}[scale=.3,rotate=45]
\filldraw[black,rounded corners=2,fill=gray!50,very thick] (0,0) rectangle (3,3);
\filldraw[black,rounded corners=2,fill=gray!20,very thick] (0,3) rectangle (1,5);
\filldraw[black,rounded corners=2,fill=gray!20,very thick] (3,2) rectangle (5,3);
\filldraw[black,rounded corners=2,fill=gray!20,very thick] (2,0) rectangle (3,-2);
\filldraw[black,rounded corners=2,fill=gray!20,very thick] (0,0) rectangle (-2,1);
\end{tikzpicture} 
&
\begin{tikzpicture}[scale=.3,rotate=45]
\filldraw[black,rounded corners=2,fill=gray!50,very thick] (0,0) rectangle (3,3);
\filldraw[black,rounded corners=2,fill=gray!20,very thick] (0,3) rectangle (2,4);
\filldraw[black,rounded corners=2,fill=gray!20,very thick] (3,3) rectangle (4,1);
\filldraw[black,rounded corners=2,fill=gray!20,very thick] (3,0) rectangle (1,-1);
\filldraw[black,rounded corners=2,fill=gray!20,very thick] (0,0) rectangle (-1,2);
\end{tikzpicture}
&
\begin{tikzpicture}[scale=.3,rotate=45]
\filldraw[black,rounded corners=2,fill=gray!50,very thick] (0,0) rectangle (1,1);
\filldraw[black,rounded corners=2,fill=gray!20,very thick] (0,1) rectangle (1,5);
\filldraw[black,rounded corners=2,fill=gray!20,very thick] (1,1) rectangle (5,0);
\filldraw[black,rounded corners=2,fill=gray!20,very thick] (0,0) rectangle (1,-4);
\filldraw[black,rounded corners=2,fill=gray!20,very thick] (0,0) rectangle (-4,1);
\end{tikzpicture}
&
\begin{tikzpicture}[scale=.3,rotate=45]
\filldraw[black,rounded corners=2,fill=gray!50,very thick] (0,0) rectangle (1,1);
\filldraw[black,rounded corners=2,fill=gray!20,very thick] (1,0) rectangle (3,2);
\filldraw[black,rounded corners=2,fill=gray!20,very thick] (0,0) rectangle (2,-2);
\filldraw[black,rounded corners=2,fill=gray!20,very thick] (0,1) rectangle (-2,-1);
\filldraw[black,rounded corners=2,fill=gray!20,very thick] (1,1) rectangle (-1,3);
\end{tikzpicture}
&
\begin{tikzpicture}[scale=.3,rotate=45]
\filldraw[black,rounded corners=2,fill=gray!50,very thick] (1,1) rectangle (2,2);
\filldraw[black,rounded corners=2,fill=gray!20,very thick] (0,1) rectangle (1,5);
\filldraw[black,rounded corners=2,fill=gray!20,very thick] (1,2) rectangle (5,3);
\filldraw[black,rounded corners=2,fill=gray!20,very thick] (2,2) rectangle (3,-2);
\filldraw[black,rounded corners=2,fill=gray!20,very thick] (2,0) rectangle (-2,1);
\end{tikzpicture} \\
\hline
\end{array}
$$
\caption{\small Five types of windmill have to be considered when $p$ is a prime number congruent to $1$ mod $4$.}
\label{5_types}
\end{figure}

The boundary cases $x=2y$ and $y=x+z$ do not happen when $p$ is prime, so we need not include them; neither do we need to worry about $x=0,y=0$ or $z=0$. The key observation is that for any given \textit{outer shape} of one of these windmills, there are \textit{either one or two} windmills that fit in it. So to each windmill, a \textit{dual} windmill can be associated -- self-dual windmills being those for which only one possible windmill has the same outer shape. This duality defines an involution on the set of windmills of a given area, which action on the dimensions $(x,y,z)$ is precisely given by the set of equations (\ref{zag_involution}). It so happens that windmills of types $2$, $3$ and $4$ result in the same equation, which is the reason why Zagier's definition distinguishes three cases and not five.

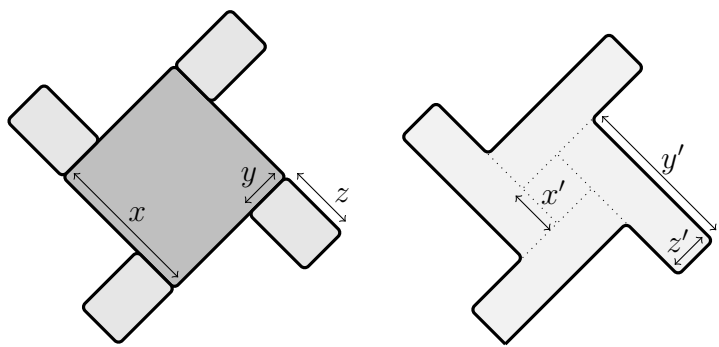
\begin{figure}[h!]
\begin{center}
\begin{tikzpicture}[scale=.7,rotate=45]
\filldraw[black,rounded corners=2,fill=gray!50,very thick] (0,0) rectangle (3,3);
\filldraw[black,rounded corners=2,fill=gray!20,very thick] (0,3) rectangle (1,4.5);
\filldraw[black,rounded corners=2,fill=gray!20,very thick] (3,2) rectangle (4.5,3);
\filldraw[black,rounded corners=2,fill=gray!20,very thick] (2,0) rectangle (3,-1.5);
\filldraw[black,rounded corners=2,fill=gray!20,very thick] (0,0) rectangle (-1.5,1);
\draw[black,<->] (.2,.1) -- (.2,2.9);
\draw[black,<->] (2.1,.2) -- (2.9,.2);
\draw[black,<->] (3.2,-.1) -- (3.2,-1.4);
\draw (.5,1.5) node {$x$};
\draw (2.5,.5) node {$y$};
\draw (3.5,-1) node {$z$};
\end{tikzpicture} 
\hspace{.1in}
\begin{tikzpicture}[scale=.65,rotate=45]
\filldraw[fill=gray!10,draw=black,rounded corners=2,very thick] (-1.5,0)-- (2,0) -- (2,-1.5) -- (3,-1.5) -- (3,2) -- (4.5,2) --(4.5,3) -- (1,3) -- (1,4.5) -- (0,4.5) -- (0,1) -- (-1.5,1) -- (-1.5,0);
\draw[black,rounded corners=2,dotted] (0,1) -- (2,1);
\draw[black,rounded corners=2,dotted] (2,0)--(2,2);
\draw[black,rounded corners=2,dotted] (3,2) -- (1,2);
\draw[black,rounded corners=2,dotted] (1,3) --(1,1);
\draw[black,<->] (.8,1) -- (.8,2);
\draw[black,<->] (2.1,-1.3) -- (2.9,-1.3);
\draw[black,<->] (3.2,1.9) -- (3.2,-1.4);
\draw (1.4,1.5) node {$x'$};
\draw (2.5,-1) node {$z'$};
\draw (3.6,.25) node {$y'$};
\end{tikzpicture} 
\end{center}
\vspace{-.1in}
\caption{\small A windmill associated to a triple $(x,y,z)$ and its outer shape. In dotted lines is its image by the Zagier involution, corresponding to the third line of (\ref{zag_involution}).}
\label{outer_shape}
\end{figure}

One can see from Figures \ref{5_types} and \ref{outer_shape} that windmills of type $1$ (resp. $2$) are paired with windmills of type $5$ (resp. $4$), whereas windmills of type $3$ are self-dual and constitute the fixed point(s) of Zagier's map; it is also quite straightforward, from considerations of area, to prove that such a fixed point occurs and is unique when $p$ is a prime congruent to $1$ mod 4. Figures \ref{fig1}, \ref{fig2}, \ref{fig3} below illustrate this for $p=17$, where $|S_p|=5$; windmills with the same outer shape are represented side by side. \medskip

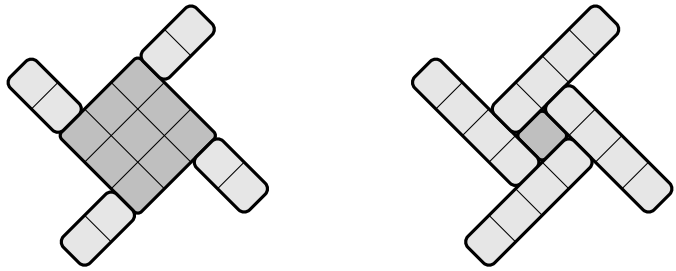
\begin{figure}[h!]
\begin{center}
\begin{tikzpicture}[scale=.5,rotate=45]
\filldraw[black,rounded corners=2,fill=gray!50,very thick] (0,0) rectangle (3,3);
\filldraw[black,rounded corners=3,fill=gray!20,very thick] (0,3) rectangle (1,5);
\filldraw[black,rounded corners=3,fill=gray!20,very thick] (3,2) rectangle (5,3);
\filldraw[black,rounded corners=3,fill=gray!20,very thick] (2,0) rectangle (3,-2);
\filldraw[black,rounded corners=3,fill=gray!20,very thick] (0,0) rectangle (-2,1);
\draw[black] (0,1)--(3,1);
\draw[black] (0,2)--(3,2);
\draw[black] (1,0)--(1,3);
\draw[black] (2,0)--(2,3);
\draw[black] (0,4)--(1,4);
\draw[black] (4,3)--(4,2);
\draw[black] (2,-1)--(3,-1);
\draw[black] (-1,0)--(-1,1);
\end{tikzpicture}
\hspace{.6in}
\begin{tikzpicture}[scale=.5,rotate=45]
\filldraw[black,rounded corners=2,fill=gray!50,very thick] (1,1) rectangle (2,2);
\filldraw[black,rounded corners=3,fill=gray!20,very thick] (0,1) rectangle (1,5);
\filldraw[black,rounded corners=3,fill=gray!20,very thick] (1,2) rectangle (5,3);
\filldraw[black,rounded corners=3,fill=gray!20,very thick] (2,2) rectangle (3,-2);
\filldraw[black,rounded corners=3,fill=gray!20,very thick] (2,0) rectangle (-2,1);
\draw[black] (0,1)--(3,1);
\draw[black] (0,2)--(3,2);
\draw[black] (1,0)--(1,3);
\draw[black] (2,0)--(2,3);
\draw[black] (0,4)--(1,4);
\draw[black] (4,3)--(4,2);
\draw[black] (2,-1)--(3,-1);
\draw[black] (-1,0)--(-1,1);
\draw[black] (0,3)--(1,3);
\draw[black] (3,3)--(3,2);
\draw[black] (2,0)--(3,0);
\draw[black] (0,0)--(0,1);
\end{tikzpicture}
\end{center}
\vspace{-.2in}
\caption{\small Windmills of type $1$ ($2y<x$) are paired with windmills of type $5$ ($y>x+z$). Above are $(3,1,2)$ and $(1,4,1)$ in $S_{17}$.}
\label{fig1}
\end{figure}

\begin{figure}[h!]
\begin{center}
\begin{tikzpicture}[scale=.5,rotate=45]
\filldraw[black,rounded corners=2,fill=gray!50,very thick] (0,0) rectangle (3,3);
\filldraw[black,rounded corners=3,fill=gray!20,very thick] (0,3) rectangle (2,4);
\filldraw[black,rounded corners=3,fill=gray!20,very thick] (3,3) rectangle (4,1);
\filldraw[black,rounded corners=3,fill=gray!20,very thick] (3,0) rectangle (1,-1);
\filldraw[black,rounded corners=3,fill=gray!20,very thick] (0,0) rectangle (-1,2);
\draw[black] (-1,1)--(3,1);
\draw[black] (0,2)--(4,2);
\draw[black] (1,0)--(1,4);
\draw[black] (2,-1)--(2,3);
\end{tikzpicture}
\hspace{.6in}
\begin{tikzpicture}[scale=.5,rotate=45]
\filldraw[black,rounded corners=2,fill=gray!50,very thick] (0,0) rectangle (1,1);
\filldraw[black,rounded corners=3,fill=gray!20,very thick] (1,0) rectangle (3,2);
\filldraw[black,rounded corners=3,fill=gray!20,very thick] (0,0) rectangle (2,-2);
\filldraw[black,rounded corners=3,fill=gray!20,very thick] (0,1) rectangle (-2,-1);
\filldraw[black,rounded corners=3,fill=gray!20,very thick] (1,1) rectangle (-1,3);
\draw[black] (1,1)--(3,1);
\draw[black] (2,0)--(2,2);
\draw[black] (1,0)--(1,-2);
\draw[black] (0,-1)--(2,-1);
\draw[black] (0,0)--(-2,0);
\draw[black] (-1,-1)--(-1,1);
\draw[black] (0,1)--(0,3);
\draw[black] (-1,2)--(1,2);
\end{tikzpicture}
\end{center}
\vspace{-.2in}
\caption{\small Windmills of type $2$ ($x/2<y<x$) are paired with windmills of type $3$ ($x<y<x+z$). Above are $(3,2,1)$ and $(1,2,2)$ in $S_{17}$.}
\label{fig2}
\end{figure}
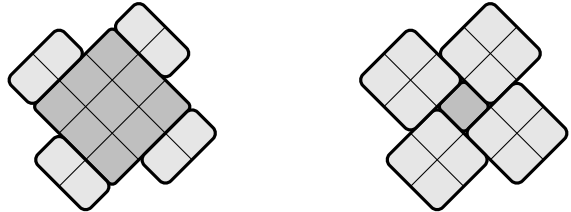

\begin{figure}[h!]
\begin{center}
\begin{tikzpicture}[scale=.5,rotate=45]
\filldraw[black,rounded corners=2,fill=gray!50,very thick] (0,0) rectangle (1,1);
\filldraw[black,rounded corners=3,fill=gray!20,very thick] (0,1) rectangle (1,5);
\filldraw[black,rounded corners=3,fill=gray!20,very thick] (1,1) rectangle (5,0);
\filldraw[black,rounded corners=3,fill=gray!20,very thick] (0,0) rectangle (1,-4);
\filldraw[black,rounded corners=3,fill=gray!20,very thick] (0,0) rectangle (-4,1);
\draw[black] (-3,0)--(-3,1);
\draw[black] (-2,0)--(-2,1);
\draw[black] (-1,0)--(-1,1);
\draw[black] (2,0)--(2,1);
\draw[black] (3,0)--(3,1);
\draw[black] (4,0)--(4,1);
\draw[black] (0,2)--(1,2);
\draw[black] (0,3)--(1,3);
\draw[black] (0,4)--(1,4);
\draw[black] (0,-1)--(1,-1);
\draw[black] (0,-2)--(1,-2);
\draw[black] (0,-3)--(1,-3);
\end{tikzpicture}
\end{center}
\vspace{-.2in}
\caption{\small $(1,1,4)$ is the fixed point of the Zagier involution on $S_{17}$.}
\label{fig3}
\end{figure}
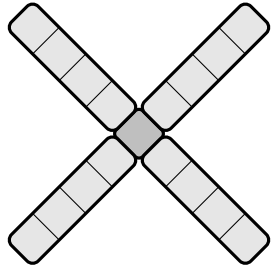

This geometric picture should help convince the \textit{bona fide} sceptical reader that Zagier's formulae (\ref{zag_involution}) indeed define an involution on $S_p$. At the same time, it should be acknowledged that checking this fact \textit{without} Spivak's windmill interpretation is more of a challenge, and requires some time and concentration, even though, admittedly, it only requires to go through a sequence of elementary steps.

\subsection{Formal verification of Zagier's proof}

Although the windmill picture is a very helpful one in order to convince an audience (and therefore a powerful observation as far as the proof as a social fact is concerned), \Coq does not need this interpretation; nor does it need this distinction of five cases to perform the necessary computations. Instead, it can simply enumerate the 9 different cases of applying (\ref{zag_involution}) twice to a triple of numbers, and discharge each of them either by proving equality or a contradiction. Doing this kind of enumeration is just as uninteresting on a computer as it is on paper, hence we automated it by writing \texttt{Ltac} tactics to do \emph{proof search}. In addition, we wrote some tactics to automate and give names to common tasks. We used \texttt{Ltac} because of our existing experience with it; as a result, the way the tactics operate does not necessarily reflect the general philosophy of \ssr (for example, the tactic \texttt{hyp\_progress} is a bookkeeping operation on a hypothesis) -- but this does not conflict with the rest of the system. The following main tactics show up in various places in our proofs:
\begin{itemize}
    \item \texttt{by\_contradiction H}: starts a proof by contradiction by doing a case analysis on the current (boolean) goal.
    In one case, the goal is automatically satisfied and thus discharged.
    This leaves the contrapositive case, which is now asserted by a hypothesis named \texttt{H}.
    The tactic inverts itself if the goal is a negated boolean, and also has a shortform \texttt{by\_contra} if the name of the hypothesis should be automatically generated.
    \item \texttt{hyp\_progress H}: takes a hypothesis of the form \texttt{X -> Y}, introduces a new goal \texttt{X} and then modifies the hypothesis \texttt{H} to only consist of \texttt{Y} in the original goal.
    \item \texttt{destruct\_boolhyp H}: automatically destructs the boolean connectives \texttt{$\|$} and \texttt{$\&\&$} in hypothesis \texttt{H} recursively, moving their components back to the goal to be individually named.
    Also tries to apply distributivity first to reduce the number of new goals.
    \item \texttt{mccontradiction}: discharges a goal whose hypotheses form an obvious contradiction, like \texttt{$H_1: a < b, H_2 : b < a$} or \texttt{$H_1: X, H_2:\; \sim\sim X$}.
    \item \texttt{mcnia}/\texttt{mclia}: uses another (otherwise unused) tactic \texttt{mctocoq} to translate MathComp-style arithmetic expressions in the context and the goal into core Coq arithmetic expressions and then applies the \texttt{micromega}~\cite{micromega} tactics \texttt{nia}/\texttt{lia} included with Coq's standard distribution to discharge the goal.
    The tactic is limited to the kinds of expressions that we see in our proofs; in the future one would probably want to use mczify~\cite{mczify} -- the only reason that we did not was that mczify was not available in the standard Windows distribution of Coq at the time we wrote our proofs.
    \item \texttt{mcsolve}: tries some basic transformations on the hypotheses (e.g. \texttt{destruct\_boolhyp}) and then tries to apply \texttt{auto}, \texttt{mcnia}, or \texttt{mccontradiction} to discharge the goal.
    \item \texttt{zag\_solve}: Our main proof search tactic.
    Among other things, introduces all hypotheses from the goal, looks for boolean if-then-else blocks and does a case analysis on their condition, destructs elements that are pattern-matched in let-expressions, and splits pair/tuple equality goals into equality goals of their constituting components.
    After these simplifications, it tries to apply mcsolve to each generated goal.
\end{itemize}

Most parts of our proof then try to quickly get to a state where one of the proof search tactics, in the most general case \texttt{zag\_solve}, can take over. \medskip

The formal verification of Zagier's proof begins by invoking the general notions about involutions, presented in Section \ref{involution_section}, by specifying that \texttt{K} is here $\mathbb{N}^3$, as Zagier's set $S_p$ contains triples of integers.
\begin{lstlisting}
Definition N3 : Type := nat * nat * nat.
Definition involutionN3 := (@involution_on [choiceType of N3]).
Definition fixedsetN3 :=(@fixedset [choiceType of N3]).
Definition involutionN3_lemma := (@involution_lemma [choiceType of N3]).
\end{lstlisting}
The next lines introduce a generic prime number $p$ and transport the MathComp finite set \texttt{`I\_p} (that corresponds to integers between $0$ and $p-1$) into an object of type \texttt{\{fset nat\}}, and triples of that type.
\begin{lstlisting}
Variable p : nat.
Variable p_prime : prime p.
Definition Ipfset : {fset nat} := [fsetval n in 'I_p].
Definition Ipf3 : {fset N3} := (Ipfset `*` Ipfset `*` Ipfset).
\end{lstlisting}
We can now define Zagier's set $S_p$ as the triples of integers $(x,y,z)$ with $x^2+4yz=p$, which we call \texttt{area} by reference to the area of a windmill (although the geometric interpretation is of no particular use here).
\begin{lstlisting}
Definition area (t:N3) : nat := (t.1.1)^2 + 4 * (t.1.2) * (t.2).
Definition S : {fset N3} := [fset t:N3 | t \in Ipf3 & (p == area t)].
\end{lstlisting}
Note that the finite set \texttt{Ipf3} is used above in order to ensure that \texttt{S} itself is finite; without this bound, it would simply not be possible to define it as an object of type \texttt{\{fset N3\}}. We will need to complement this with a lemma that proves that this bound actually follows from the area condition -- so that this set \texttt{S} truly is the same as $S_p$. This lemma is called \texttt{bound\_Sp}. \medskip

The last things we need to define are the two maps introduced by Zagier from the set $S_p$ to itself. We define them as maps from $\mathbb{N}^3$ to $\mathbb{N}^3$ and will verify that they induce `involutions on \texttt{S}' in the sense explained in Section \ref{involution_section}.
\begin{lstlisting}
Definition zig (t : N3) :N3 := (t.1.1, t.2, t.1.2).

Definition zag (t:N3) : N3 := match t with (x,y,z) =>
     if y >= (x + z) then (x + 2 * z, z, y - (x + z))
else if (2 * y) >= x then (2 * y - x, y,z + x - y)
                       else (x - 2 * y, z + x - y, y) end.
\end{lstlisting}
The theorems that follow these definitions establish, one by one, the different requirements of Zagier's proof; that is, everything that was left to the reader, such as the fact that \texttt{zig} and \texttt{zag} are involutions on \texttt{S}. These lemmas, for instance, are called \texttt{zig\_involution} and \texttt{zag\_involution}. Here is what the proof of the second one (arguably the most challenging claim, in this proof, to check with paper an pen) looks like:
\begin{lstlisting}
Lemma zag_involution: involutionN3 S zag.
Proof.
rewrite /involution_on; split; move => [[x y] z].
 - rewrite !inE /area /zag /Ipfset /= /area /zag => hin.
   destruct_boolhyp hin => hx hy hz /eqP hp.
   have harea_p := area_p hp.
   zag_solve.
 - rewrite !inE /zag => htS; destruct_boolhyp htS => hx hy hz /eqP hp.
   have harea_p := area_p hp.
   zag_solve.
Qed.
\end{lstlisting}
Other prerequisites are written similarly, that is, with the help of \MC libraries, the \ssr syntax, and the tactics presented above that make part of the analysis automatic. Once this preliminary work has been done, the lines of the final proof can essentially be interlaced with Zagier's original sentence:

\begin{lstlisting}
Theorem Fermat_Zagier : p %% 4 = 1 -> exists a b :nat, p = a^2 + b^2.
Proof.
move /modulo_ex => [k hk].
\end{lstlisting}
\vspace{-.05in}
\hfill `The involution on the finite set \texttt{S} defined by \texttt{zag}'
\vspace{-.1in}
\begin{lstlisting}
have h_zag_invol:=zag_involution.
\end{lstlisting}
\hfill `has exactly one fixed point,' 
\vspace{-.1in}
\begin{lstlisting}
have h_zag_fix_card:(#|`(fixedsetN3 S zag)|) = 1.
   - by rewrite (zag_fixed_point hk); first by apply: cardfs1.
\end{lstlisting}
\hfill `so \texttt{|`S|} is odd,'
\vspace{-.1in}
\begin{lstlisting}
have h_S_odd: odd(#|`S|).
   by rewrite -(involutionN3_lemma h_zag_invol) h_zag_fix_card.
\end{lstlisting}
\hfill `and the involution defined by \texttt{zig}.' 
\vspace{-.1in}
\begin{lstlisting}
have h_zig_invol:= zig_involution.
\end{lstlisting}
\hfill `also has a fixed point.' 
\vspace{-.1in}
\begin{lstlisting}
have [t htzigfix]: exists t:N3, t \in (fixedsetN3 S zig).
  by apply odd_existence; rewrite (involutionN3_lemma h_zig_invol).
by apply (zig_solution htzigfix).
Qed.
\end{lstlisting}

\subsection{Formal verification of Theorem \ref{Jackson_thm} }

The following argument, under the assumption that $p \equiv 3 \ \rm{mod} \ 8$, provides a short proof of Theorem~\ref{Jackson_thm}.

\begin{proof}[Jackson's one-sentence proof]
The involution on the finite set
\begin{equation}\label{set_Sp_jack}
S_p := \{ (x,y,z)\in \mathbb{N}^3 \ | \ x^2+2yz=p \}
\end{equation}
defined by
\begin{equation}\label{jack_involution}
    (x,y,z) \mapsto
    \left\{
    \begin{array}{ll}
        (x-2y, z+2x-2y, y) & \text{if } 2y < x  \\
        (2y-x, y, 2x-2y+z) &  \text{if } x < 2y < 2x+z \\
        (3x-2y+2z, 2x-y+2z,-2x+2y-z) & \text{if } 2x+z < 2y < 3x+2z \\
        (-3x+2y-2z, -2x+2y-z,2x-y+2z) & \text{if } 3x+2z < 2y < 4x+4z \\
        (x+2z, z, -2x+y-2z) & \text{if } 4x+4z < 2y
    \end{array}
    \right.
\end{equation}
has exactly one fixed point, so $|S_p|$ is odd, and the involution defined by 
\begin{equation}\label{zig_involution_jack}
    (x,y,z) \mapsto (x,z,y).
\end{equation}
also has a fixed point.
\end{proof}

As one can see, the structure of the argument is exactly the same as Zagier's one-sentence proof, the only differences being that the involution \eqref{jack_involution} is not the same as \eqref{zag_involution}; it operates on a different finite set, distinguishes five cases instead of three, and is not an involution under the same assumption (indeed, the assumptions that  $p \equiv 1 \ \rm{mod} \ 4$ or  $p \equiv 3 \ \rm{mod} \ 8$ are mutually exclusive). There does not seem to be a geometric interpretation for this involution. \medskip

Our method works the same on this example, with very minimal adaptation. One only needs to provide a proof of simple facts on squares modulo 8 that can be checked by direct case analysis. One also needs to be particularly careful in writing down the definition of involution \eqref{jack_involution}; not only to avoid typos (which goes without saying), but to make sure that smaller terms are always subtracted from larger ones under the assumption of the corresponding line. Due to the definition of subtraction on type \texttt{nat}, writing $3x -2y +2z$ on line 3 does not define the right map -- but it does with the proper ordering $3x + 2z - 2y$.

\section{David Christopher's proof}\label{Partition_section}

An alternative proof of Fermat's theorem was recently given by David Christopher \cite{DavidChristopher2016}, which shares with Zagier's proof the appeal to involutions, used in both ways. The sets on which these involutions are defined have a different (arguably more elementary) structure than Zagier's set $S_p$. We present this method below; first in the classical sense, then our formal verification of it.

\subsection{General presentation of the argument}

\begin{figure}[h!]
\begin{center}
\begin{tikzpicture}[scale=.65,rotate=45]
\filldraw[black,rounded corners=2,fill=gray!50,very thick] (0,0) rectangle (2,4);
\filldraw[black,rounded corners=3,fill=gray!20,very thick] (2,0) rectangle (5,3);
\draw[black,<->] (-.4,0)--(-.4,4);
\draw[black] (-1,2) node (A1) {$a_1$};
\draw[black,<->] (5.4,0)--(5.4,3);
\draw[black] (6,1.5) node (A2) {$a_2$};
\draw[black,<->] (0,-.4)--(1.95,-.4);
\draw[black] (1,-1) node (F1) {$f_1$};
\draw[black,<->] (2.05,-.4)--(5,-.4);
\draw[black] (3.5,-1) node (F2) {$f_2$};
\draw[black] (1,0)--(1,4);
\draw[black] (3,0)--(3,3);
\draw[black] (4,0)--(4,3);
\draw[black] (0,1)--(5,1);
\draw[black] (0,2)--(5,2);
\draw[black] (0,3)--(2,3);
\draw[black,dashed] (0,0)--(4,4);
\end{tikzpicture}
\hspace{.2in}
\begin{tikzpicture}[scale=.65, rotate=45]
\filldraw[black,rounded corners=2,fill=gray!50,very thick] (0,0) rectangle (3,5);
\filldraw[black,rounded corners=3,fill=gray!20,very thick] (3,0) rectangle (4,2);
\draw[black,<->] (-.4,0)--(-.4,5);
\draw[black] (-1.4,2.5) node (A1') {$f_1+f_2$};
\draw[black,<->] (4.4,0)--(4.4,2);
\draw[black] (5,1) node (A2') {$f_1$};
\draw[black,<->] (0,-.4)--(2.95,-.4);
\draw[black] (1.5,-1) node (F1) {$a_2$};
\draw[black,<->] (3.05,-.4)--(4,-.4);
\draw[black] (3.7,-1.2) node (F2) {$a_1-a_2$};
\draw[black] (1,0)--(1,5);
\draw[black] (2,0)--(2,5);
\draw[black] (0,1)--(4,1);
\draw[black] (0,2)--(3,2);
\draw[black] (0,3)--(3,3);
\draw[black] (0,4)--(3,4);
\draw[black,dashed] (0,0)--(4,4);
\end{tikzpicture}
\end{center}
\vspace{-.2in}
\caption{\small Young diagram corresponding to an integer partition of $p=17$ with two parts, and its conjugate.}
\label{fig4}
\end{figure}
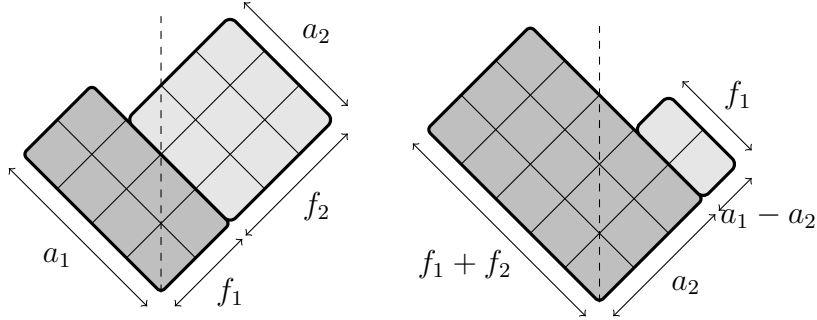

The main set of interest in David Christopher's proof encodes integer partitions of $p$, which parts take exactly \textit{two} values, $a_1$ and $a_2$, with $a_1 >a_2$. 
\begin{equation}
    \mathscr{P}_2^p := \{ (a_1, f_1, a_2, f_2 ) \in \mathbb{N}^4 \ | \ a_1>a_2, \ p=a_1 f_1 + a_2 f_2 \}.
\end{equation}
An element of $\mathscr{P}_2^p$ can be represented by a Young diagram with two steps, as illustrated on Figure \ref{fig4}. The set of such diagrams is stable by conjugation, which induces the following involution on $\mathscr{P}_2^p$:
\begin{equation}
    \tau : (a_1, f_1, a_2, f_2)   \mapsto (f_1+f_2, a_2, f_1, a_1-a_2)
\end{equation}

It is straightforward to check that $\tau$ has exactly one fixed point. Indeed, the conditions for being a fixed point are 
\begin{equation}
    a_1=f_1+f_2 \ \ \& \ \ f_1=a_2
\end{equation}
so that expressing $a_1, a_2$ in terms of $f_1, f_2$ gives
$p=f_1 (f_1 + 2 f_2)$. It follows that $f_1=1$, which in turns determines $f_2 = \frac{p-1}{2}$. Note that this part of the argument only relies on $p$ being an \textit{odd} prime number. The conclusion is that $|\mathscr{P}_2^p|$ is odd. \medskip

Another natural idea is to swap each pair $a_1, f_1$ and $a_2,f_2$, but as such it does not define an involution on $\mathscr{P}_{2}^{p}$: for instance, if $f_1=f_2$ then the outcome is not in $\mathscr{P}_{2}^{p}$ as it is not a partition with two parts, but one. However, this idea works on well-defined subsets. We consider the following partition of $\mathscr{P}_{2}^p$:
\begin{equation}
    \mathscr{P}_{2}^p = \mathscr{P}_{2,>}^p \sqcup \mathscr{P}_{2,=}^p \sqcup \mathscr{P}_{2,<}^p
\end{equation}
where
\begin{align}
\mathscr{P}_{2,<}^p & := \{ (a_1, f_1, a_2, f_2 ) \in \mathscr{P}_{2}^p \ | \ f_1 < f_2 \} \label{P2-} \\
\mathscr{P}_{2,=}^p & := \{ (a_1, f_1, a_2, f_2 ) \in \mathscr{P}_{2}^p \ | \ f_1 = f_2 \} \label{P2=} \\
\mathscr{P}_{2,>}^p & :=\{ (a_1, f_1, a_2, f_2 ) \in \mathscr{P}_{2}^p \ | \ f_1 > f_2 \} \label{P2+}
\end{align}
and we make a few straightforward observations.

\noindent First, concerning (\ref{P2-}), we note that the double-swapping $$(a_1,f_1,a_2,f_2) \mapsto (f_2,a_2,f_1,a_1)$$
defines an involution of $\mathscr{P}_{2,<}^p$ that does not have a fixed point (such a fixed point would imply $p=2 a_1 f_1$, which is absurd). We conclude, invoking Lemma \ref{FLI}, that $|\mathscr{P}_{2,<}^p|$ is even. \medskip

\noindent The next observation is that the set $\mathscr{P}_{2,=}^p$ defined by (\ref{P2=}) can be enumerated directly; as $p$ is prime, an element of $\mathscr{P}_{2,=}^p$ must be such that $f_1=f_2=1$. As there are exactly $(p-1)/2$ ways of writing $p=a_1+a_2$ with $a_1>a_2$, it follows that 
$$|\mathscr{P}_{2,=}^p| = \frac{p-1}{2},$$
which is even. A partial conclusion at that point is that $|\mathscr{P}_{2,>}^p|$ is odd.\medskip

\noindent The last observation is that the simple swapping
$$(a_1,f_1,a_2,f_2) \mapsto (f_1,a_1,f_2,a_2)$$
defines an involution on $\mathscr{P}_{2,>}^p$, a finite set of odd cardinality by the previous arguments. By Lemma \ref{FLI}, it must have a fixed point, which implies $a_1=f_1, a_2=f_2$, and therefore
$p=a_1^2+a_2^2$, a sum of two squares. This concludes David Christopher's proof.

\subsection{Formal verification of David Christopher's proof}

A formal verification of this second proof can be found in the third part of the master file \texttt{all\_Zagier\_project.v}, or equivalently \texttt{proof\_B\_partition.v} after compiling the preliminary file \texttt{lemmata.v}. It begins with the following transparent definitions; clearly, the partitions of $p$ that are considered involve quadruples of integers, so that \texttt{K} is here $\N^4$.
\begin{lstlisting}
Definition N4 : Type := (nat * nat) * (nat * nat).
Definition Ip4 : {fset N4} := ((Ip `*` Ip) `*` (Ip `*` Ip)).
Definition involutionN4 := (@involution_on [choiceType of N4]).
Definition fixedsetN4 := (@fixedset [choiceType of N4]).
Definition involutionN4_lemma := (@involution_lemma [choiceType of N4]).
\end{lstlisting}
The quantity $a_1 f_1 + a_2 f_2$ is called \texttt{Y\_area}, referring to the area of a Young diagram.
\begin{lstlisting}
Definition Y_area (yd : N4) : nat := yd.1.1 * yd.1.2 + yd.2.1 * yd.2.2.
\end{lstlisting}
Finally, here are the definitions of the sets considered above, properly defined as objects of type \texttt{\{ fset N4 \}} thanks to an \textit{a priori} bound.
\begin{lstlisting}
Definition P2p :=[fset yd:N4 | (yd \in Ip4)&&((Y_area yd == p)&&(yd.1.1 > yd.2.1))].
  Definition P2p_l :=[fset yd:N4 | (yd \in P2p) && (yd.1.2 < yd.2.2)].
  Definition P2p_e :=[fset yd:N4 | (yd \in P2p) && (yd.1.2 == yd.2.2)].
  Definition P2p_g :=[fset yd:N4 | (yd \in P2p) && (yd.1.2 > yd.2.2)].
\end{lstlisting}
The formal proof then follows the argument explained above in a rather straightforward way. The general lemma on involutions and the ad hoc tactics like \texttt{mcnia} and \texttt{zag\_solve} written for Zagier's proof are used again, to the same effect. \medskip 

The final proof, written in this way, is longer than the one based on Zagier's argument, and this for two reasons: obviously, it requires to introduce more objects (sets and involutions) and to go through more steps; but also, there is one argument of a different kind needed to enumerate the middle set $\mathscr{P}_{2,=}^p$. From the point of view of formal verification, this is the only difference compared to Zagier's proof, which only relied on involutions. In order to perform this enumeration, we first define an injective function that generates quadruples from natural numbers as follows:
$$q(n) := (p-(n+1), 1, n+1, 1)$$
We then prove that $\mathscr{P}_{2,=}$ is equal to the set generated by mapping $q$ over the set of natural numbers less than $\lfloor{(p-1)/2}\rfloor$, which MathComp expresses as \texttt{I\_((p-1)\%/2)}. The cardinality of that set is of course known from the libraries.

\newpage
\section*{Conclusion}

We have formally verified two short combinatorial proofs of Fermat's sum of two squares theorem (Theorem \ref{Fermat_thm}), as well as a proof of another related result (Theorem \ref{Jackson_thm}), using the \Coq proof assistant. It was to be expected that the final result would not exactly match the length of Zagier's `one-sentence proof' when presented informally. However, we feel that these formal proofs are not outrageously long either, considering the level of exactness they warrant -- and they certainly could be further shortened if needed. More importantly, the feeling that a formal proof exactly matches the main mental steps required by the original argument, and is not something intrinsically different - but only more precise - is very satisfying. \medskip

There are other generalizations of Zagier's proof, relying on different sets and involutions, and the same technique could also be applied to formally verify these proofs beyond the example of Theorem \ref{Jackson_thm}.
The only practical issue is one of computational resources needed for proof search: the more involved the involution is, the more cases have to be considered, and the more complex some of these cases become.
We have optimized our proof search tactics to some extent to avoid exploring goal transformations that eventually fail.
However, even in the optimal case, non-linear inequalities will eventually have to be discharged by \texttt{nia}, which itself searches for positivstellensatz refutations.
Thus, as the equations get more complex, the majority of Coq's computing time and memory allocation is spent on those searches.  \medskip

For both authors, this project was a first encounter with the \MC libraries, and for one of them, a first serious endeavour with the \Coq proof assistant altogether. It would have taken a lot more time and effort without the recurrent help and detailed explanations of experts in the community; especially, one can only advise new users in a similar situation to express their concerns online as soon as they encounter a difficulty. It is also a somewhat dull but necessary task to read the documentation of each library very carefully to avoid issues or resolve them quickly. In particular, one should be careful to not only import the right libraries but also to \textit{open the right scopes in the right order}. For the sake of a (real-life) example: in order to define and use objects of type \texttt{ \{fset K \}}, it is necessary to first write
\begin{lstlisting}
Open Scope fset_scope.
\end{lstlisting}
However, one of the side-effects of this scope is to overwrite the sign \texttt{+} usually used for addition, so that it is not possible anymore to write expressions like \texttt{a+b} with \texttt{a b : nat} right after opening \texttt{fset\_scope}. One should \textit{re-open} the \texttt{nat\_scope} first:
\begin{lstlisting}
Open Scope fset_scope.
Open Scope nat_scope.
\end{lstlisting}

Another important caveat is that the Search function, useful as it is, does not necessarily yield all relationships between concepts, in particular in cases where one concept is defined in terms of another. For example, the divisibility predicate \texttt{dvdn d m} is defined using the modulo operation (as \texttt{modn m d == 0}). Writing \texttt{Search \_ modn dvdn.} does not immediately provide access to this information; instead, one needs to look directly at the definition of \texttt{dvdn}. \medskip

The above aspects are easy enough to work with, once noticed. But subtleties of that sort are perhaps the most time-consuming aspect when discovering the libraries. The sooner they are resolved, the sooner one can actually work on relevant aspects of the code itself.

\end{document}